\documentclass[a4paper]{article}

\usepackage{INTERSPEECH2022}
\usepackage{multirow}
\usepackage{color}
\usepackage{epstopdf}

\title{Enhanced exemplar autoencoder with cycle consistency loss in any-to-one voice conversion}
\name{Weida Liang, Lantian Li, Wenqiang Du, Dong Wang\thanks{
This work was supported by the National Natural Science
Foundation of China under Grant No.62171250,
and also the Tsinghua-SPD Bank Joint Research Center for Digital Finance Technologies.
Dong Wang is the corresponding author.}}
\address{
  Center for Speech and Language Technologies, Tsinghua University}
\email{\{liangwd,lilt,duwq\}@cslt.org, wangdong99@mails.tsinghua.edu.cn}

\begin{document}

\maketitle
\begin{abstract}

Recent research showed that an autoencoder trained with speech of a single speaker, called exemplar autoencoder (eAE), 
can be used for any-to-one voice conversion (VC).
Compared to large-scale many-to-many models such as AutoVC, the eAE model is 
easy and fast in training, and may recover more details of the target speaker.

To ensure VC quality, the latent code should represent and only represent content information.
However, this is not easy to attain for eAE as it is unaware of any speaker variation in model training. 
To tackle the problem, we propose a simple yet effective approach based on a cycle consistency loss.
Specifically, we train eAEs of multiple speakers with a shared encoder, and meanwhile encourage
the speech reconstructed from any speaker-specific decoder to get a consistent latent code as the original speech 
when cycled back and encoded again.
Experiments conducted on the AISHELL-3 corpus showed that this new approach improved the baseline eAE consistently.
The source code and examples are available at the project page: \emph{http://project.cslt.org/}.

\end{abstract}

\section{Introduction}

The purpose of voice conversion (VC) is to transform a speech waveform to make it sound like spoken by another person
while preserve the linguistic content.
Early studies mainly focused on learning a frame-level one-to-one mapping function, by employing parallel 
data~\cite{stylianou1998continuous,chen2014voice,nakashika2014voice,takashima2013exemplar}.
This approach is obviously costly, and the application is limited to one-to-one conversion. 

Modern voice conversion methods are based on large-scale training with non-parallel data. The basic idea is to
learn a disentanglement model that can separate content and speaker information in speech signals, 
and then perform conversion by picking up content information from the source speaker and speaker information from the target speaker. 
Most of the approaches are based on the encoder-decoder architecture,
where the encoder produces the content code, and the decoder
uses them to synthesize the target speech by referring to a new speaker code. 
Representative models include PPG~\cite{xie2016kl,sun2016phonetic}, 
VAE~\cite{luong2021many}, CVAE~\cite{hsu2016voice,kameoka2019acvae,qian2019autovc,qian2020unsupervised},
VQVAE~\cite{van2017neural,wu2020vqvc}, and AdaIN-VC~\cite{chou2019one}.

In spite of the promising performance, these large-scale models require a big amount of data. 
Although this is not a serious problem for rich-resource 
languages, for low-resource language, collecting content/speaker labelled data 
does impose problems. Moreover, 
training models with large datasets is not economic in 
energy consumption. Finally, using a single speaker code to modulate the shared decoder 
might be insufficient to represent details of the speaker's trait, 
for example prosody~\cite{qian2020unsupervised,qian2021global}.
The pay-off of the large-scale training, of course, is 
the ability to perform (nearly) any-to-any conversion. For example, with the AutoVC~\cite{qian2019autovc} model
trained on more than 100 speakers in VCTK~\cite{veaux2017cstr}, speech of any speaker
can be converted to any target speaker, with perhaps just one enrollment utterance.

In some cases, however, the any-to-any capacity is not the most required. 
For example, I like voice from Albert Einstein, and wish all 
the lectures being taught in his voice. In this scenario, any-to-one is
sufficient and perhaps more suitable: it is easy to train, 
and may recover more details of the target speaker as the decoder is dedicated to that person. 
The exemplar autoencoder (eAE)~\cite{deng2020unsupervised} is such a model.
It is a vanilla autoencoder (AE) trained with a bunch of utterances (typically tens of minutes) 
of a particular speaker. The authors argued that 
if human speech is content-dominated, i.e., content varies more than speaker trait, 
the eAE model can convert speech of any speaker to the target speaker (speaker for training). 
However, content-domination is dubious in practice; 
even if it is held, there is no guarantee that the conversion is of anything perfect or good.
A basic reason is that eAE tends to produce information entangled code, as the model
does not see any speaker variation when it is trained.

\begin{figure}[htb!]
\centering
\includegraphics[width=0.85\linewidth]{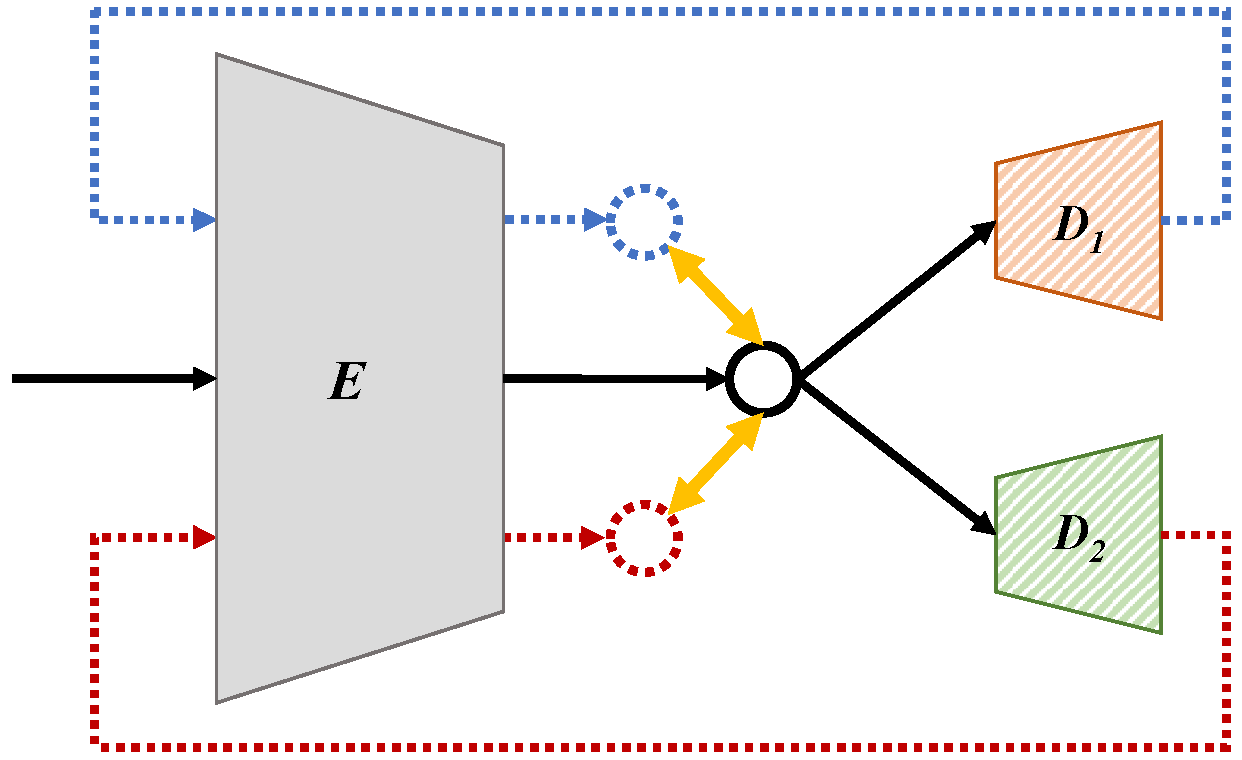}
\caption{Concept of cycle consistency loss.}
\label{fig:concept}
\end{figure}

In this paper, we propose to solve the entanglement problem of eAE
by a cycle consistency loss. As shown in Fig.~\ref{fig:concept}: Firstly, 
we train a multi-head eAE with data of multiple speakers, 
where the encoder is shared while the decoders are speaker-specific (blue and red dash lines). 
Secondly, we encourage the reconstructed speech close to the original speech in the \emph{code} space,
no matter which decoder is used to perform the reconstruction (yellow arrows).
We will show that this simple cycle consistency loss leads to 
much more disentangled code, even with just two training speakers.  

\begin{figure*}[htb!]
\centering
\includegraphics[width=0.92\linewidth]{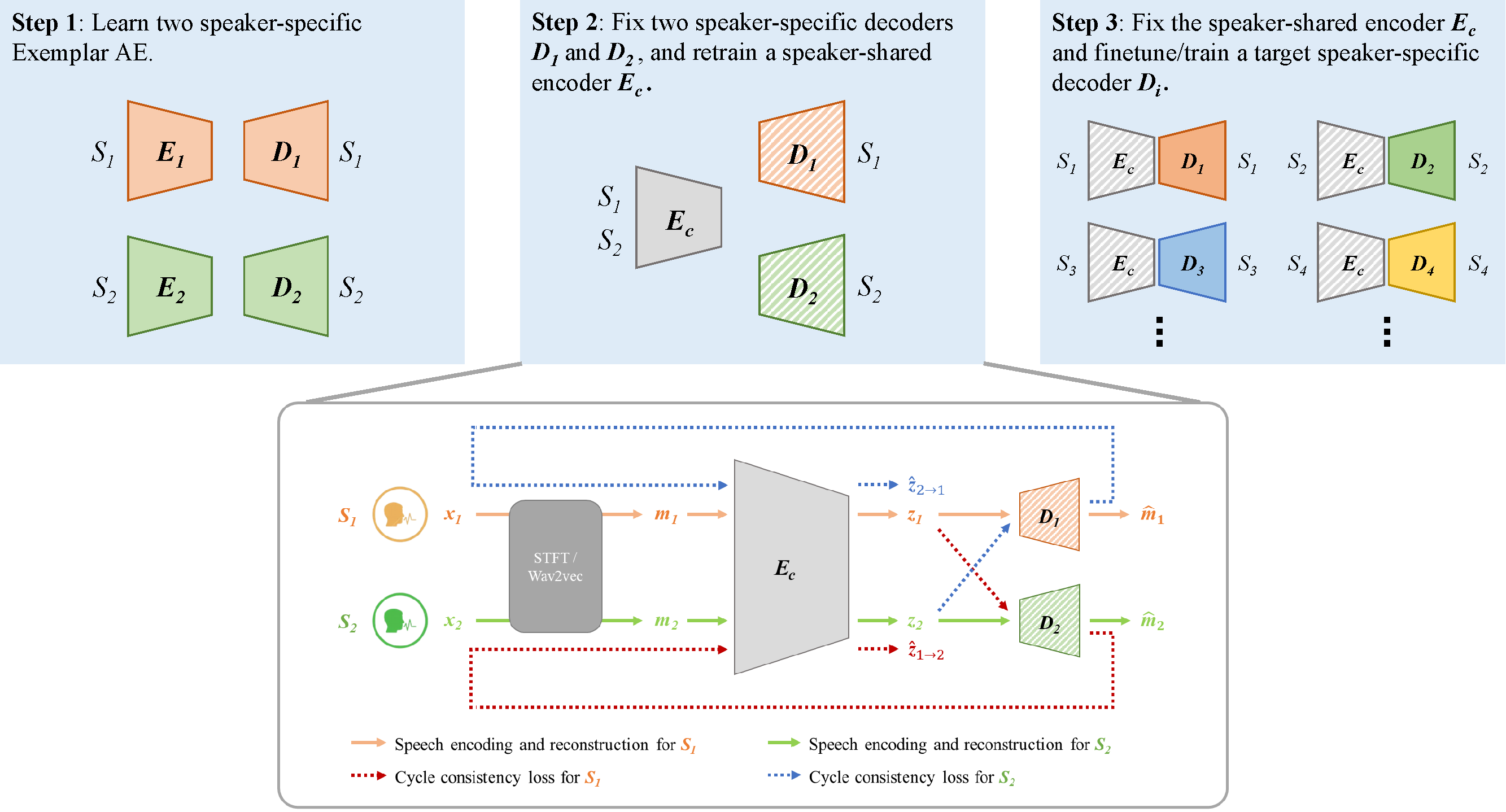}
\caption{The framework of enhanced exemplar autoencoder.}
\label{fig:ae}
\end{figure*}

\section{Related Works}
\label{sec:rel}

There are numerous studies in VC. 
Readers are referred to recent review papers for more technical discussions~\cite{mohammadi2017overview,sisman2020overview}.
Focusing on any-to-one VC, the PPG + synthesizer architecture~\cite{xie2016kl,sun2016phonetic,yeh2018rhythm} is
a powerful candidate. The shortage is that an ASR system is required, which is impossible for 
low-resource languages. Using pre-trained models (e.g., Wav2Vec~\cite{lin2021s2vc,huang2021any} or HuBERT~\cite{kreuk2021textless}) 
to substitute ASR systems alleviates the difficulty, however the generated codes are not necessarily speaker-invariant.
We will show in the experiments that even with a strong Wav2Vec encoder, the cycle consistency loss still contributes. 

Information disentanglement has been investigated by multiple authors 
in VC, mostly by penalizing mutual information (MI) between the content code and the speaker code, including
adversarial training~\cite{ocal2019adversarially,wang2021adversarially} and inter-code MI penalty~\cite{wang2021vqmivc}.
These MI regularization methods require extra classifier/discriminator, and suffer from unstable training. 
Moreover, with a very limited number of speakers (2 speakers in our case), computing the MI is not very meaningful. 

The idea of cycle consistency loss has been known in CycleGAN~\cite{zhu2017unpaired,lee2020many}
and CycleVAE~\cite{tobing2019non,matsubara2021high}. 
Our proposal is fundamentally different from these prior work in that our cycle is code-to-code, 
while the cycle in CycleGAN and CycleVAE is data-to-data. 
We will show by experiments that this difference is not trivial.


\section{Methodology}
\label{sec:method}

\subsection{Cycle consistency loss for eAE}

Before any theoretical discussion for the property and advantage,
let's first describe how the cycle consistency loss is implemented. 
To make the presentation clear, we use two speeches $x_1$, $x_2$ 
from two speakers $S_1$, $S_2$ to describe the process. The 
diagram is shown in Fig.~\ref{fig:ae}.

\begin{itemize}
\item \textbf{Step 1. Train two separate speaker-specific eAEs.} 
First transform $x_1$ and $x_2$ to Mel spectrum  $m_1$ and $m_2$ via STFT,
and train two separate eAEs using $m_1$ and $m_2$ following the process in~\cite{deng2020unsupervised}, 
denoted by \{$E_1$,$D_1$\} and \{$E_2$,$D_2$\} respectively.


\item \textbf{Step 2. Fix the decoders $\bm{D_1}$ and $\bm{D_2}$, and retrain a speaker-shared encoder $\bm{E_c}$ with cycle consistency loss.}

(1) Reinitialize a new encoder $E_c$, and then forward $m_1$ and $m_2$ to $E_c$, obtaining the content codes $z_1$ and $z_2$.

(2) Forward $z_1$ and $z_2$ to $D_1$ and $D_2$ respectively to generate $\hat{m}_1$ and $\hat{m}_2$. The reconstruction loss is computed as follows:

\begin{equation}
\label{eq:rec}
\mathcal{L}_{rec} =  ||{\hat{m}_1}- {m_1}||^2 + ||{\hat{m}_2}- {m_2}||^2,
\end{equation}
\noindent where $||\cdot||$ denotes the $\ell_2$-norm. See the yellow and green solid lines in Fig.~\ref{fig:ae}.

(3) Forward $z_1$ and $z_2$ to  $D_2$ and $D_1$ respectively (note the code and decoder do not match).
The decoding output is then cycled back  and fed to the encoder $E_c$, producing the recycled code  $\hat{z}_{1\rightarrow 2}$ and $\hat{z}_{2 \rightarrow 1}$.
The cycle consistency loss is computed as follows:

\begin{equation}
\label{eq:cyc}
\mathcal{L}_{cyc} = ||{\hat{z}_{1 \rightarrow 2}}- {z_1}||^2 + ||\hat{z}_{2 \rightarrow 1}- {z_2}||^2,
\end{equation}
\noindent See the red and blue dash lines in Fig.~\ref{fig:ae}.

(4) The final loss combines the reconstruction loss in Eq.~(\ref{eq:rec}) and the cycle loss in Eq.~(\ref{eq:cyc}):

\begin{equation}
\mathcal{L} =  \mathcal{L}_{rec}+\alpha * \mathcal{L}_{cyc},
\end{equation}
\noindent where $\alpha$ is a hyper-parameter to control the ratio of $\mathcal{L}_{rec}$ and $\mathcal{L}_{cyc}$.
In our experiment, we empirically set $\alpha$ to 10.

\item \textbf{Step 3. Fix the speaker-shared encoder $\bm{E_c}$ and finetune existing decoders $\bm{D_1}$, $\bm{D_2}$ or train new decoders $\bm{D_3}$, $\bm{D_4}$ for new target speakers.}

\end{itemize}

During conversion, pass the source speech $x$ to $E_c$, and choose a decoder $D_i$ 
corresponding to the target speaker to perform decoding, producing reconstructed spectrum $\hat{m}$. 
A WaveNet vocoder~\cite{oord2016wavenet} is then employed to synthesize the converted speech $\hat{x}$.

\subsection{Theoretical analysis}
\label{sec:anal}

This section presents theoretical analysis for the cycle consistency loss. 
We start from analyzing why the vanilla eAE tends to fail in practice. 

\subsubsection{eAE is fragile}

We first recite the theoretical augment for eAE presented in~\cite{deng2020unsupervised}. 
Define $s$ and $w$ as the speaker factor and content factor respectively,
and let $x=f(s,w)$ denote the speech with content $w$ and spoken by speaker $s$, 
where $f$ is the \emph{true} generating process.

In~\cite{deng2020unsupervised}, the authors argued that in the Mel spectrum space, human speech
is content-dominated, i.e., speech signals of the same word but spoken by different speakers 
are more similar than those of different words spoken by the same speaker. Formally speaking:

\vspace{-1mm}
\begin{equation}
  D_{Mel}(f(s_1,w_0),f(s_2,w_0)) \leq D_{Mel}(f(s_1,w_0),f(s_2,w)) \ \  \forall w,
\end{equation}
\noindent where $D_{Mel}$ denotes distance in Mel spectrum.  This assumption is equal to say:

\vspace{-1mm}
\begin{equation}
\label{eq:eae}
  f(s_2, w_0) = \arg\min_{x \in M} D(x,f(s_1,w_0)),
\end{equation}
\noindent where $M = \{f(s_2,w): w \in W\}$.

Suppose that the eAE model is trained by speech of $s_2$, it essentially learns the 
manifold of $s_2$. Therefore, for any speech $f(s_1,w_0)$, the eAE function tends to project 
it to the manifold $M$, which is the RHS of Eq.(\ref{eq:eae}). This leads to 
the converted version $f(s_2,w_0)$.

The argument above, however, seems over strong. Firstly, although it is generally 
true that content variation is more significant than speaker variation at the word level, 
while at the phone level, speaker change could be more drastic. 
This means that the phone content will be changed by the eAE projection due to speaker variation.
Unfortunately, small change on phones may cause serious impact in perception. 
More qualitative evidences please refer to\footnote{http://project.cslt.org/}.
Moreover, the manifold assumption heavily relies on a tight information bottleneck (IB) 
on the code, which is not easy to set.

\subsubsection{Cycle consistency loss works}

The fragility of eAE mentioned above is essentially caused by the content-speaker 
information entangled in the code; this is in turn caused by the single-speaker training
scheme, which makes eAE unaware of any speaker variation. To solve the problem, one may
resort to either a pre-trained front such as Wav2Vec~\cite{lin2021s2vc} or large-scale multi-speaker training as 
AutoVC~\cite{qian2019autovc}. 
However, with these `heavy' solution, the main advantage of eAE is lost. 

The cycle consistency loss is a light-weighted solution. We analyze the asymptotic behavior at the 
optimum point, i.e., when $L_{cyc}=0$. In this scenario, we have:

\vspace{-1mm}
\begin{equation}
\label{eq:ohmygod}
\hat{z}_{i \rightarrow j} = z_i \ \ \forall i \ \ \forall j. 
\end{equation}
\noindent This in essence implies that on the \emph{reconstructed} speech, 
$\hat{z}$ is independent of $D_j$. 
If we further assume that all the decoders can
perfectly recover the speech of their corresponding speakers, 
then the reconstructed speech and original speech are exactly the same.
According to Eq.(\ref{eq:ohmygod}), this implies that the code $z$ of \emph{true} speech is independent of $D_j$.
If the number of speakers is abundant and $\{D_j\}$ can approximate the entire 
population, $z$ will be truly speaker independent. Surprisingly, we will show 
in the experiments that by the cycle consistency loss, 
2 different-gender speakers are sufficient to train an enhanced eAE that produces 
fairly independent code $z$.

Note that in the above derivation, we do not impose any constraint on IB. However, 
IB is still crucial, otherwise $D_j$ could be identical for all $j$. In this case,
$z$ is independent of $D_j$ but not the speaker, and VC becomes impossible.

\section{Experiments}
\label{sec:exp}

In this section, the proposed enhanced eAE is applied to any-to-one voice conversion task.

\subsection{Data}

We use speech data from the AISHELL-3 dataset~\cite{shi2020aishell} to construct the training and test sets, as shown in Table~\ref{tab:data}.
All the speech signals are formatted with 16kHz sampling rate and 16-bits precision.
No overlap in speakers exists between the training and test sets.

\begin{table}[htbp]
\caption{Data profile}
\vspace{-2mm}
\label{tab:data}
\centering
\scalebox{0.8}{
\begin{tabular}{clll}
\toprule
\textbf{Set}   & \textbf{\# of Spks}     &  \textbf{Utters per Spk}  &  \textbf{Duration per Spk} \\
\midrule
Train             & 4  (2 Female, 2 Male)   &  $\sim$400                &  $\sim$25 mins     \\
Test              & 6  (3 Female, 3 Male)   &  $\sim$250                &  $\sim$15 mins     \\
\bottomrule
\end{tabular}}
\vspace{-2mm}
\end{table}

\subsection{Model}

All the eAE models were implemented following~\cite{deng2020unsupervised} with the source code\footnote{https://github.com/dunbar12138/Audiovisual-Synthesis}.

\noindent \textbf{Preprocessing:} The training speech is firstly clipped into segments with 1.6s in length,
each being further transformed to a spectrogram by STFT with window size of 800 and hop size of 200. 
The spectrogram is then transformed to 80$\times$128 Mel-spectrogram.

\noindent \textbf{Encoder:} The Mel-spectrogram is fed to the encoder which consists of three 1D convolutional layers,
each followed by a normalization and ReLU activation. The kernel size of each layer is 5 and the number of channels is 512.
The output of the convolutional layers is then fed to two bidirectional LSTM layers with cell dimensions of 32,
resulting in a 32-dimensional content code.

\noindent \textbf{Decoder:} The content code is passed through the decoder.
It is firstly up-sampled to the original time resolution and then input to one 512-channel LSTM layer and
three 512-channel 1D convolutional layers with a kernel size of 5. Each layer is accompanied with batch normalization and ReLU activation.
Finally, the output is fed into two 1,024-channel LSTM layers and a fully connected layer that produces 80-dim Mel-spectrograms.

\noindent \textbf{Vocoder:} We train a WaveNet vocoder~\cite{oord2016wavenet} to convert Mel-spectrograms to speech signals.


\subsection{Metrics}

Four metrics are used for quantitative evaluation, including goodness of pronunciation (GOP), character error rate (CER),
MOSNet score and speaker classification accuracy (SCA). GOP and MOSNet primarily evaluate the quality of the generation, 
CER mostly focuses on intelligibility, and SCA is more related to resemblance to the target speaker.  

The Kaldi toolkit~\cite{povey2011kaldi} is used to compute CER and GOP. The pre-trained model in~\cite{lo2019mosnet} is used to predict the MOSNet score.
For SCA test, we train a speaker classification model based on the x-vector structure~\cite{snyder2018x} 
with 400 background speakers from AISHELL-1 dataset~\cite{bu2017aishell} plus the target speakers from the training set. SCA is computed 
as the accuracy of the converted speech classified to the target speaker.

\subsection{Main results}
\label{sec:main}

We use two speakers in the training set to train the eAE with and without cycle consistency loss.
The 6 speakers in test set (3 males and 3 females) are used to perform two groups of test: 
same-gender (SG) and cross-gender (CG). Results are reported in Table~\ref{tab:basic}.

Firstly, it can be observed that with the cycle consistency loss, 
the performance on all the evaluation metrics has been improved consistently and substantially.
Secondly, the improvement is more significant in the cross-gender test, leading to performance 
very similar to that in the same-gender test. This result indicates that the eAE model indeed 
suffers from large speaker variation, and applying the cycle consistency loss can alleviate the problem 
to a large extent. 

\begin{table}[ht]
\caption{Comparison between eAEs with/without cycle consistency loss. SG and CG denote the same-gender and
cross-gender tests respectively.}
\vspace{-2mm}
\label{tab:basic}
\centering
\resizebox{\columnwidth}{!}{%
\begin{tabular}{llcccc}
\toprule
           &           & \textbf{GOP} ($\uparrow$)& \textbf{CER(\%)} ($\downarrow$)& \textbf{MOSNet} ($\uparrow$) & \textbf{SCA(\%)} ($\uparrow$)\\
\midrule
 \multirow{2}*{eAE}       & SG              & 1.489        & 19.29            & 2.712           & 81.85            \\
                          & CG              & 1.368        & 21.19            & 2.668           & 80.00            \\
\midrule
 \multirow{2}*{eAE + Cycle} & SG              & 1.605        & 14.27            & 2.786           & 85.00            \\
                          & CG              & 1.589        & 14.19            & 2.778           & 85.45            \\
\bottomrule
\end{tabular}}
\vspace{-2mm}
\end{table}

\subsection{Generalization to new target speakers}

In this test, we firstly train an eAE with cycle consistency loss as in the previous experiment, 
and then fix the encoder $E_c$ and train decoders for 6 new speakers selected from AISHELL-3.
The same test data in the test set are used to perform test on these new target speakers. 
For comparison, we also train 6 individual vanilla eAEs for the same 6 speakers. 
The results are reported in Table~\ref{tab:unseen}.

It can be seen that the pre-trained encoder $E_c$ can be effectively adapted to the new speakers, and bring 
significant performance improvement.
This further verifies our conjecture that with the cycle consistency loss,
the speaker-shared encoder can produce more disentangled content codes.

\begin{table}[htbp]
\caption{Performance on new target speakers.}
\vspace{-2mm}
\label{tab:unseen}
\centering
\scalebox{0.92}{
\begin{tabular}{lccc}
\toprule
 & \textbf{GOP} ($\uparrow$)& \textbf{CER(\%)} ($\downarrow$)& \textbf{MOSNet} ($\uparrow$) \\ 
\midrule
 eAE             & 1.439           & 20.86             & 2.718            \\
 eAE + Cycle     & \textbf{1.539}  & \textbf{15.23}    & \textbf{2.760}   \\
\bottomrule
\end{tabular}}
\vspace{-2mm}
\end{table}

\subsection{Ablation study}

We perform a couple of ablation studies to understand the behavior of the cycle consistency loss. 
The same target speaker and test speech as in Section~\ref{sec:main} is used to perform the study. 
For simplicity, we only report the results on the cross-gender group, where the cycle consistency loss 
contributes the best. 
The overall results are reported in Table~\ref{tab:ablation}. For an easy comparison, 
results of the eAE baseline and eAE + Cycle baseline are reproduced, shown as Model 1 and Model 2 respectively.

\begin{table}[ht]
\caption{Results of ablation study.}
\vspace{-1mm}
\label{tab:ablation}
\centering
\scalebox{0.76}{
\begin{tabular}{llcccccc}
\toprule
 \textbf{No.} & \textbf{Model} & \textbf{\# Spks} & \textbf{GOP} & \textbf{CER(\%)} &  \textbf{MOSNet} & \textbf{SCA(\%)} \\
\midrule
  1&eAE               & 1                & 1.368          & 21.19             & 2.768            & 80.00             \\
  2&eAE + Cycle       & 2                & 1.589          & 14.19             & 2.778            & 85.45    \\
\midrule
  3&eAE + Cycle       & 4                & 1.593          & 14.03             & 2.737            & 85.10             \\
\midrule
  4&eAE + En-Share    & 2                & 1.378          & 21.28             & 2.689            & 80.40             \\
\midrule
  5&eAE + Data Cycle  &2                & 1.513          & 18.56             & 2.724            & 82.80           \\
\midrule
  6&eAE/W2V           &2                 & 1.612          & 11.88             & 2.795            & 89.25             \\
  7&eAE/W2V + Cycle   &2                & 1.713         & 10.73             & 2.823            &  89.60    \\ 
\bottomrule
\end{tabular}}
\vspace{-1mm}
\end{table}

\subsubsection{More training speakers}

To test if more training speakers provides additional gains, we use 4 speakers to train an enhanced eAE model.
The results are shown as Model 3. By comparing with Model 2, we can see that using two additional speakers does not offer clear 
advantage, indicating that perhaps 2 speakers are sufficient for the cycle consistency loss to train a reasonable encoder.

\subsubsection{Encoder sharing or cycle loss?}

One may argue the improvement obtained by the enhanced eAE could be due to the shared encoder, 
rather than the cycle consistency loss. To response this argument, 
Model 4 reports the results with encoder sharing only but no cycle loss. 
Comparing to eAE (Model 1) and advanced eAE (Model 2), 
we see that only sharing the encoder does not provide much advantage,
confirming that cycle consistency loss is the essence of our proposal.

\subsubsection{Code cycle and data cycle}

The cycle consistency loss presented in the paper follows a code cycle, by $z_1 \rightarrow \hat{z}_{1\rightarrow 2}$.
It is also possible to design a data cycle, following the path 
$x_1\rightarrow \hat{x}_{1 \rightarrow 2} \rightarrow \hat{x}_{1 \rightarrow 2 \rightarrow 1} $.
This is the way that CycleVAE follows~\cite{tobing2019non,matsubara2021high}. We implemented the 
data cycle scheme as Model 5. Comparing to eAE (Model 1) and advanced eAE (Model 2), 
it can be observed that the data cycle indeed improves eAE, but much less significant than the code cycle.

\subsubsection{Work with powerful frontend}

Inspired by~\cite{huang2021any}, we use a pre-trained VQW2V model\footnote{https://dl.fbaipublicfiles.com/fairseq/wav2vec/wav2vec\_small.pt}
as a more powerful front, to test if the cycle loss still contributes. 
The results are reported as Model 6 and Model 7.
It can be seen that the W2V front provides clear improvement over the Mel spectrum front (Model 6 vs. Model 1 and Model 7 vs. Model 2).
Even using a strong front, training with cycle loss still offers additional gains (Model 6 vs Model 7).

\section{Conclusion}
\label{sec:con}

In this paper, we proposed an enhanced exemplar autoencoder for any-to-one voice conversion.
The core design is a cycle consistency loss, which enforces the content code of the reconstructed speech 
close to the original speech, no matter by whose decoder decodes the speech.
We demonstrated theoretically and empirically that the proposed technique can significantly purify the content code,
and produce better performance in complex VC tasks, such as cross-gender conversion.
In future, we would like to conduct deep investigation on the behavior of the cycle consistency loss,
and apply the loss to other VC models.


\bibliographystyle{IEEEtran}
\bibliography{mybib}

\end{document}